\documentclass[twocolumn,a4paper]{article}

\usepackage{xspace}
\usepackage{graphicx}
\usepackage[english]{babel}
\usepackage{times}
\usepackage[scaled=0.92]{helvet}

\usepackage{mathrsfs}
\usepackage{bm,theorem,amsmath,amssymb}
\usepackage{url}
\usepackage[a4paper]{geometry}

\advance\textheight by 4cm
\advance\topmargin -2cm
\advance\textwidth by 3cm
\advance\oddsidemargin by -1.5cm
\advance\evensidemargin by -1.5cm

\usepackage{paralist}

\newcommand{\lst}[2]{$#1_0$,~$#1_1$, $\dots\,$,~$#1_{#2-1}$}
\newcommand{\lstl}[2]{$#1_0$,~$#1_1$, $\dots\,$,~$#1_{#2}$}

\def\..{\,\mathpunct{\ldotp\ldotp}} % Middle stuff for intervals. Usage: \..

\title{Quasi-Succinct Indices}

\author{Sebastiano Vigna\\
Dipartimento di Informatica, Universit\`{a} degli Studi di
Milano, Italy}

\begin{document}
\bibliographystyle{abbrv}

\maketitle

\begin{abstract}
Compressed inverted indices in use today are based on the idea of gap
compression: documents pointers are stored in increasing order, and the
gaps between successive document pointers are stored using suitable codes
which represent smaller gaps using less bits. Additional data such as
counts and positions is stored using similar techniques. A large body of
research has been built in the last 30 years around gap compression,
including theoretical modeling of the gap distribution, specialized
instantaneous codes suitable for gap encoding, and \textit{ad hoc} document
reorderings which increase the efficiency of instantaneous codes. This paper proposes to represent an
index using a different architecture based on quasi-succinct
representation of monotone sequences. We show that, besides being
theoretically elegant and simple, the new index provides expected
constant-time operations and, in practice, significant performance
improvements on conjunctive, phrasal and proximity queries.\end{abstract}

\section{Introduction}

An \emph{inverted index} over a collection of documents contains, for
each term of the collection, the set of documents in which the term appears and
additional information such as the number of occurrences of the term within each
document, and possibly their positions. Inverted indices form the backbone of
all modern search engines, and the existence of large document collections (typically, the web)
has made the construction of efficient inverted indices ever more important. 

Compression of inverted indices saves disk space, but more importantly also
reduces disk and main memory accesses~\cite{BuCICGERA}, resulting in faster
evaluation. We refer the reader to the book by Manning, Raghavan and
Sch{\"u}tze~\cite{MRSIIR} and to the very complete and recent survey by Zobel
and Moffat~\cite{ZoMIFTSE} for a thorough bibliography on the subject.

Two main complementary techniques are at the basis of index compression:
\emph{instantaneous codes} provide storage for integers that is proportional to
the size of the integer (e.g., smaller numbers use less bits); \emph{gap
encoding} turns lists of increasing integers (for instance, the
monotonically increasing list of numbers of documents in which a term appear)
into lists of small integers, the \emph{gaps} between successive values
(e.g., the difference). The two techniques, combined, make it possible to store
inverted indices in highly compressed form. Instantaneous codes are also
instrumental in storing in little space information such as the number of
documents in which each term appears.

Since inverted indices are so important for search engines, it is not surprising
that a large amount of research has studied how to maximize either the speed or
the compression ratio of gap-encoded indices. Depending on the application,
compression or speed may be considered more important, and different solutions
propose different tradeoffs.

In this paper, we describe a new type of compressed index that does not use
gaps. Rather, we carefully engineer and tailor to the needs of a search engine a
well-known \emph{quasi-succinct} representation for monotone sequences
proposed by Peter Elias~\cite{EliESRCASF}.\footnote{Incidentally, Elias also invented some of the
most efficient codes for gap compression~\cite{EliUCSRI}.} We explain how to
code every part of the index by exploiting the bijection between sequence of integers
and their \emph{prefix sums}, and we provide details about the physical storage
of our format.

Our new index is theoretically attractive: it guarantees to code the information
in the index close to its \emph{information-theoretical lower bound}, and
provides on average constant-time access to any piece of information stored in
the index, including searching for elements larger than a given value (a
fundamental operation for computing list intersections quickly). This happens by
means of a very simple addressing mechanism based on a linear list
of forward pointers. Moreover, sequential scanning can be performed using a
very small number of logical operation per element. We believe it is particularly attractive
for in-memory or memory-mapped indices, in which the cost of disk access is not dominant.

To corroborate our findings, in the last part of the paper, we index the TREC
GOV2 collection and a collection or 130 million page of the \texttt{.uk}
web\footnote{We remark that TREC GOV2 is publicly available, and that the latter
collection is available from the author.} with different type of encodings, such
as $\delta$ and Golomb. We show that, while not able to beat gaps coded with
Golomb codes, our index compresses better than $\gamma$/$\delta$ codes or
variable-length byte.
 
We then compare a prototype Java implementation of our index against MG4J and
Lucene, two publicly available Java engine based, and Zettair, a C search
engine. MG4J has been set up to use $\gamma$/$\delta$ codes, whereas Lucene and
Zettair use variable-length byte codes. We get a full
confirmation of the good theoretical properties of our index, with excellent
timings for conjunctive, phrasal and proximity queries. We also
provide some evidence that for pointers list our index is competitive with the
Kamikaze implementation of PForDelta codes~\cite{ZHNSSRCCC}.

The quasi-succinct indices described in this paper are the default indices
used by MG4J from version
5.0.\footnote{\texttt{http://mg4j.di.unimi.it/}}

\section{Related work}

The basis of the current compression techniques for inverted indices is
\emph{gap encoding}, developed at the start of the '90s~\cite{BMNDCFTRS}. Gap
encoding made it possible to store a positional inverted index in space often \emph{smaller}
than the compressed document collection. Gaps (differences between
contiguous document pointers in the posting list) have to be encoded using instantaneous codes that
use shorter codewords for smaller integers, and previous research in information theory provided~$\gamma$, $\delta$~\cite{EliUCSRI} and
Golomb~\cite{GolRLE} codes, which achieve excellent compression. Moreover, a
wealth of alternative codes have been developed in the last 30 years.\footnote{Alternative approaches, such as \emph{interpolative
coding}~\cite{MoSECIFC}, have been proposed to code some part of an index, but
they lack the direct-access and skipping features that are necessary for fast
query resolution.}

When speed is important, however, such codes are rather slow to decode: in
practice, often implementation use the folklore \emph{variable-length byte
code} (e.g., the open-source search engine Lucene, as well as
Zettair). Recent research has developed a number of \emph{word-aligned}
codes~(e.g., \cite{AnMIICUWABC}) that encode in a single machine word several
integers, providing high-speed decoding and good compression.
In~\cite{ZHNSSRCCC}, the author tailor their PForDelta code to the behavior of
modern super-scalar CPUs and their caches.

More specialized techniques tackle specific problems, studying in great detail
the behaviour of each part of the index: for instance, \cite{YDSCTPWI} studies
in great detail the compression of positional information.

Another line of research studies the renumberings of the documents that generate
smaller gaps. This phenomenon is known as \emph{clustering}~\cite{MoSECIFC}, and
can be induced by choosing a suitable numbering for the
documents~\cite{BlBTCBSRDI,SPOADIECWSEI,BlBICDR}.

As indices became larger, a form of \emph{self-indexing}~\cite{MoZSIIFFTR}
became necessary to compute quickly the intersection of lists of documents, an
operation that is at the basis of the computation of conjunctive Boolean
queries, proximity queries and phrasal queries.

The techniques used in this paper are based on a seminal paper by
Elias~\cite{EliESRCASF}, which is a precursor of \emph{succinct data structures}
for indexed sets~\cite{RRSSID}. We do use some of the knowledge developed by the
algorithmic community working on succinct data structures, albeit in practice
the theoretical encodings developed there, which concentrate on 
attaining asymptotically optimal speed using $o(n)$
additional bits, where $n$ is the optimal size for the data structure, 
have presently too high constant costs to be competitive in real applications
with methods using $O(n)$ additional bits.

We remark that the literature on the subject is actually immense, and impossible
to recap in this section. The references above should be considered mostly as
pointers. We refer the reader again to~\cite{MRSIIR,ZoMIFTSE} for
a complete historical overview.

\section{Definitions}

In this paper we discuss the indexing problem for a collection of documents. We
give definitions from scratch as we will need to discuss formally the index
content.

Each document is represented by a number, called \emph{document
pointer}, starting from zero. Each document $d$ has a length $\ell$, and is
formed by a sequence of terms \lst {t}{\ell}. For each document and each term,
the \emph{count} specifies how many times a term appears in the sequence
forming the document. The \emph{frequency} is the number of documents in which a
term appears (i.e., the number of documents for which the count is not zero). 
The \emph{occurrency} of a term is the number of occurrences of the term in the
whole collection, that is, the sum of the counts of the term over all documents.

The \emph{posting list} for a term is the (monotonically increasing) list of
documents where the term appears. With each document we associate also the
(nonzero) count of the term in the document, and the (monotonically increasing)
list of \emph{positions} (numbered from zero) at which the term appears in the
given document.

The \emph{unary code} associates
with the natural number $n\geq 0$ the codeword $0^n1$. The \emph{negated unary
code} associates with the natural number $n\geq 0$ the codeword $1^n0$.

A \emph{bit array} of length $n$ is a sequence of bits \lst bn. We sometime view
such an array as a \emph{stream}: we assume that there is an implicit pointer,
and that I/O operations such as reading unary codes are performed by
scanning the array and updating the implicit pointer accordingly.

\section{Quasi-Succinct Representation of Monotone Sequences}
\label{sec:qsrep}
\begin{figure}
\centering
\includegraphics[scale=.9]{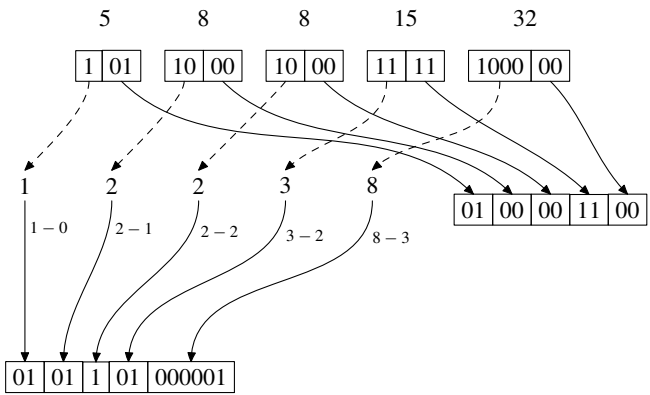}
\caption{\label{fig:repr}A simple example of the quasi-succinct encoding
from~\cite{EliESRCASF}. We consider the list $5$, $8$, $8$, $15$, $32$ with upper bound $36$, so
$\ell=\lfloor\log(36/5)\rfloor=2$. On the right, the lower $\ell$ bits of
all elements are concatenated to form the lower-bits array. On the left,
the gap of the values of the upper bits are stored sequentially in unary code
in the upper-bits array.}
\end{figure}

In this section we give a detailed description of the 
\emph{high bits/low bits} representation of a
monotone sequence proposed by Elias~\cite{EliESRCASF}.
% \footnote{As
% reported by Elias, Fano had independently had discovered the same
% representation~\cite{FanNBRISM}.} 
We assume to have a
monotonically increasing sequence of $n>0$ natural numbers
\[
0\leq x_0\leq x_1 \leq \cdots \leq x_{n-2} \leq x_{n-1}\leq u,
\]
where $u>0$ is any upper bound on the last value.\footnote{If $u=0$, the list
is entirely made of zeroes, and its content is just defined by $n$.} The choice
$u=x_{n-1}$ is of course possible (and optimal), but storing explicitly $x_{n-1}$ might be costly, and a suitable value for $u$ might be known from external information, as we will see shortly.
We will represent such a sequence in two bit arrays as follows:
\begin{itemize}
  \item the lower $\ell=\max\{\,0,\lfloor \log (u/n)\rfloor\,\}$ bits of each
  $x_i$ are stored explicitly and contiguously in the \emph{lower-bits
  array};\footnote{Actually, Elias discusses just the case in which $u+1$ and $n+1$ are powers of two, but extending his definitions is an easy exercise.}
  \item the upper bits are stored in the \emph{upper-bits array} as a
  sequence of unary-coded gaps.
\end{itemize}
In Figure~\ref{fig:repr} we show an example. Note that we code the gaps between
the values of the upper bits, that is, $\bigl\lfloor
x_i/2^\ell\bigr\rfloor-\bigl\lfloor x_{i-1}/2^\ell\bigr\rfloor$ (with the
convention $x_{-1}=0$).

The interesting property of this representation is that it uses at most 
$2+\lceil\log(u/n)\rceil$ bits per element: this can be easily seen from the
fact that each unary code uses one stop bit, and each other written bit increases the
value of the upper bits by $2^\ell$: clearly, this cannot happen more than
$\bigl\lfloor x_{n-1}/2^\ell\bigr\rfloor$ times. But
\begin{equation}
\label{eq:bound}
\biggl\lfloor\frac {x_{n-1}}{2^\ell}\biggl\rfloor\leq\biggl\lfloor\frac
u{2^\ell}\biggl\rfloor\leq\frac u{2^\ell} = \frac u{2^{\max\{\,0,\lfloor \log (u/n)\rfloor\,\}}} \leq 2n.
\end{equation}
Thus, we write at most $n$ ones and $2n$ zeroes, which implies our statement as
$\lceil\log(u/n)\rceil=\lfloor\log(u/n)\rfloor+1$ unless $u/n$ is a power of
two, but in that case~(\ref{eq:bound}) actually ends with $\leq
n$, so the statement is still true. 

Since the information-theoretical lower
bound for a monotone list of $n$ elements in a universe of $u$ element is
\[
\Biggl\lceil\log{u+n\choose n}\Biggr\rceil\approx n
\log\biggl(\frac{u+n}n\biggr)
\]
we see that the representation is close to succinct:
indeed, Elias proves in detail that this representation is very close to
the optimal representation (less than half a bit per element away). Thus, while
it does not strictly classify as a \emph{succinct} representation, it can be
safely called a \emph{quasi-succinct} representation.\footnote{Actually, the
representation is one of the ingredients of sophisticated, modern succinct
data structures that attain the information-theoretical bound~\cite{RRSSID}.}

To recover $x_i$ from the representation, we perform $i$ unary-code reads
in the upper-bits array, getting to position $p$: the
value of the upper bits of $x_i$ is then exactly $p-i$; the lower $\ell$ bits
can be extracted with a random access, as they are located at position $i\ell$ in the
lower-bits array.

We now observe that, assuming to have a
fictitious element $x_{-1}=0$, we can equivalently see the list \lst xn as a list of natural
numbers by computing gaps: \[ a_0 =x_0-x_{-1}, a_1= x_1-x_0,
\cdots,a_{n-1}=x_{n-1}-x_{n-2}. \] Conversely, given a list \lst an of natural numbers we can
consider the list of \emph{prefix sums} $s_k=\sum_{i=0}^{k-1} a_i$ for $0\leq k \leq n$. The
two operations give a bijective correspondence between monotone
sequences\footnote{Note that sequences of prefix sums contain an additional
element $s_0=0$ that is not part of the bijection.} bounded by $u$ and
lists of natural numbers of the same length whose sum is bounded by
$u$.\footnote{The same bijection is used normally to code monotone sequences
using gaps, but we intend to to the opposite.} Thus, we can represent using the high bits/low bits
presentation either monotonically increasing sequences, or generic lists of integers.\footnote{Prefix sums have indeed several applications in compression, for instance to the storage of XML documents~\cite{DRRCPS}.}

The quasi-succinct representation above has a number of useful
properties that make it quite advantageous over gap-encoded sequences:
\begin{itemize}
  \item The distribution of the document gaps is irrelevant: there is no
  code to choose, because the lower bits are stored explicitly in a fixed-width format,
  and the representation of the upper bits, being made by $n$ ones and at
  most $2n$ zeroes, is a perfect candidate for the unary code. 
  \item Compression is guaranteed irrespective of gaps being well distributed
  (e.g., because of correlation between the content of consecutive document) or
  not. In particular,
  renumbering documents in a way that improves retrieval speed (e.g., to ease
  early termination) will not affect the index size.
  \item Scanning sequentially the list using a longword buffer requires to
  perform just a unary read and using few shifts for each element.
  \item In general, the high bits/low bits representation
  concentrates the difficulty of searching and skipping on a simple bit array of unary codes containing
  $n$ ones and at most $2n$ zeroes. We can devise extremely fast, practical
  \textit{ad hoc} techniques that exploit this information.
\end{itemize}

Actually, Elias's original paper suggests the most obvious solution for quick
(on average, constant-time) reading of a sequence of unary codes: we store
\emph{forward pointers} to the positions (inside the upper-bits array) that
one would reach after $kq$ unary-code reads, $k\geq0$, where $q$ is a fixed \emph{quantum} (in other words, we record the position immediately after the one of index
$kq-1$ in the bit array).

Retrieving $x_i$ now can be done by simulating $q\lfloor i/q\rfloor$ unary reads
using a forward pointer, and completing sequentially with $i\bmod q<q$
unary-code reads. On average, by~(\ref{eq:bound}), the sequential part will read at most $3q$
bits.\footnote{This problem is essentially (i.e., modulo an off-by-one) the
\emph{selection} problem for which much more sophisticated solutions, starting
with Clarke's~\cite{ClaCPT}, have in the last years shown that constant-time
access can be obtained using $o(n)$ additional bits instead of the $O(n)$ bits
proposed by Elias, but such solutions, while asymptotically optimal,
have very high constant costs. Nonetheless, there is a
large body of theoretical and practical knowledge that has been accumulated in
the last 20 years about selection, and we will use some of the products of that
research to read multiple unary codes quickly in the upper-bits array.} Smaller
values of $q$ yield less reads and use more space.

\smallskip\noindent\textbf{Skipping.}  A more interesting property, for our
purposes, is that by storing \emph{skip pointers} to positions reached after
\emph{negated} unary-code reads of the upper bits it is possible to perform \emph{skipping},
that is, to find very quickly, given a bound $b$, the smallest $x_i\geq b$. This
operation is fundamental in search engines as it is the base for quick list
intersection.\footnote{Elias describes a slightly different analogous operation,
by which he finds the largest $x_i\leq b$; the operation involves moving
\emph{backwards} in the bit array, something that we prefer to avoid for
efficiency. Note that this is again essentially equivalent to \emph{predecessor
search}, a basic problem in fast retrieval on sets of integers for which very
strong theoretical results are known in the RAM model~\cite{PaTTSTOPS}.}

\begin{figure*}
\centering
\includegraphics{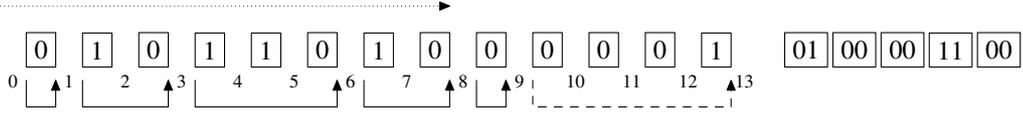}
\caption{\label{fig:skip}An example of skipping based on
the sequence shown in Figure~\ref{fig:repr}. On the left we have the
upper-bits array, and on the right the lower-bits array. We want to skip to the first item larger than or
equal to 22, so since $\ell=2$ we have to perform $\lfloor 22/2^2\rfloor=5$ negated unary-code reads (the continuous arrows), getting to position $9$, so we are
positioned in the middle of the unary code associated with the element of index
$9-5=4$. Then we perform a unary-code read (the dashed arrow), which returns
$3$, so we know that the upper bits of the current element (of index 4) are
$3+5=8$. Since the block of lower bits of index $4$ is zero, we return $32$. If we had
at our disposal a skip pointer for $q=4$ (the dotted arrow), we could have
skipped the first four negated unary-code reads. Note that in general more than
one unary-code read might be necessary after reading the negated unary codes.}
\end{figure*}

To see why this is possible, note that by definition in the upper-bits array the
unary code corresponding to the smallest $x_i\geq b$ must terminate \emph{after}
$\bigl\lfloor b/2^\ell\bigr\rfloor$ zeroes. We could thus perform $\bigl\lfloor
b/ 2^\ell\bigr\rfloor$ negated unary-code reads, getting to position $p$, and
knowing that there are exactly $p-\bigl\lfloor b/2^\ell\bigr\rfloor$ ones and
$\bigl\lfloor b/ 2^\ell\bigr\rfloor$ zeroes to our left (i.e., we are in the
middle of the unary code for $x_{p-\bigl\lfloor b/2^\ell\bigr\rfloor}$). From
here, we complete the search exhaustively, that is, we actually compute the
values of the elements of the list (by reading unary codes and retrieving the
suitable lower bits) and compare them with $b$, as clearly the element we are
searching for cannot be represented earlier in the upper-bits array. An example
is shown in Figure~\ref{fig:skip}.

By setting up an array of \emph{skip pointers} analogously to the previous
case (i.e., forward pointers), the reading of negated unary-codes can be perform
quickly. Note, however, that in general without further assumptions it is not possible to bound the number of bits read
during the $\bigl\lfloor b/2^\ell\bigr\rfloor\bmod q$ negated unary-code reads
that must be performed after following a skip pointer, as there could be few
zeroes (actually, even none) in the bit array. Nonetheless, if a linear lower bound on the number of zeroes
in the bit array is known, it can be used to show that skipping is performed in
constant time on average.

\smallskip\noindent\textbf{Strictly monotone sequences.} In case the sequence
\lst xn to be represented is \emph{strictly} monotone (or, equivalently, the
$a_i$'s are nonzero), it is possible to reduce the space usage by storing the
sequence $x_i-i$ using the upper bound $u-n$. Retrieval happens in the same
way---one just has to adjust the retrieved value for the $i$-th element by adding $i$. This mechanism was
already noted by Elias~\cite{EliBRMS} (more generally for $k$-spaced sequences,
$k>0$), but it is important to remark that under this representation \emph{the
algorithm for skipping will no longer work}. This happens because $x_i$ is
actually represented as $x_i-i$, so skipping 
$\bigl\lfloor b/2^\ell\bigr\rfloor$ negated
unary codes could move us arbitrarily \emph{after} the element we would like to
reach.

\section{Sequences as a Ranked Characteristic Functions}

In some cases, the quasi-succinct representation we described is not very
efficient in term of space: this happens, for instance, for very dense
sequences. There is however an alternate representation for strictly monotone
sequences with skipping: we simply store a list of $u$ bits in which bit $k$ is
set if $k$ is part of the list \lst xn. This is equivalent to storing the
list in gap-compressed form by writing in unary the gaps $x_i-x_{i-1}-1$, and
guarantees by definition that no more than $u$ bits will be used.

Skipping in such a representation is actually trivial: given the bound $b$, we
read a unary code starting at position $b$. The new position $x_i$ is such that
$x_i$ is the smallest element satisfying $x_i\geq b$. The only problem is that
at this point we will have lost track of the index $i$.

To solve this problem, we take a dual approach to that of the previous section
and store a simple \emph{ranking} structure: for each position $kq$, where
$q$ is the quantum, we store the number of ones to the left. After a skip, we
simply rank the current position $x_i$ by first reading the precomputed number
of ones before $\lfloor x_i /q\rfloor$, and then then computing the number of
ones in the at most $q$ remaining bits.

\section{Representing an Inverted Index}
\label{sec:repr}
We now discuss how the quasi-succinct representation presented in the
previous section can be used to represent the posting list of a term. We defer
to the next section a detailed discussion of the data-storage format.

\noindent\textbf{Pointers.} Document pointers form a strictly monotone
increasing sequence. We store them using the standard representation (i.e.,
not the specialized version for strictly monotone sequences), so to be able to
store skip pointers, as skipping is a frequent and useful operation (e.g., during the resolution of conjunctive Boolean queries or phrasal queries),
whereas random access to document pointers is not in general
necessary.\footnote{Nothing prevents from storing both kind of pointers. The increase in size of the index would be unnoticeable.} The upper bound is the number of documents $N$ minus one, and the
number of elements of the list is $f$, the frequency.

We remark that the apparent loss of compression due to the necessity of using
the standard representation (to make skipping possible) turns actually into an
advantage: if the last pointer in the list is equal to $\alpha N$, with
$0\leq\alpha<1$, since $N\geq f$, we can write $N=df+r$ with $d>0$ and $0\leq
r<f$, and then we have
\begin{equation}
\label{eq:bound2}
\biggl\lfloor\frac {\alpha N}{2^\ell}\biggl\rfloor =
\biggl\lfloor\frac{\alpha (df+r)}{2^{\lfloor \log ((df+r)/f)\rfloor}}\biggl\rfloor\geq
\biggl\lfloor\frac{\alpha (df+r)}{2^{\lfloor \log d\rfloor}}\biggl\rfloor
%\geq\biggl\lfloor \alpha f + \frac{\alpha r}d\biggl\rfloor
\geq \alpha f.
\end{equation}
In other words, the slight redundancy guarantees that there are at least
$\alpha f$ zeroes in the upper-bits array: if $\alpha \approx 1$, we can
thus guarantee that on average skipping can be performed in a constant number of
steps, as, on average, reading a one implies reading at least a zero, too (and
viceversa). Since we write forward pointers only for lists with $f\geq q$,
under realistic assumptions on $q$ in practice $\alpha $ is close to $1$.

Finally, even in pathological cases (i.e., a every uneven distribution of the
zeroes in the list), one every $2^\ell\leq N/f$ bits must necessarily be
zero, as the list is strictly monotone. Thus, terms with dense posting lists
must have frequent zeroes independently of the considerations above.

Note that if 
\[
f + \bigl\lfloor N / 2^\ell\bigr\rfloor + f\ell > N
\]
then the representation above uses more than $N$ bits (in practice, this
happens when $f\gtrsim N/3$). In this case, we switch to a ranked characteristic
function. Since there are at most two zeroes for each one in the bitmap, it
is easy to check that all operations can still be performed in average constant
time.

\noindent\textbf{Counts.} Counts are strictly positive numbers, and can
be stored using the representation for strictly monotone sequences to
increase compression. In this case the upper bound is the occurrency
of the term, and the number of elements is again the frequency.

\noindent\textbf{Positions.} The format for positions is the trickiest one.
Consider, for the $i$-th document pointer in the inverted list for term
$t$ with count $c_i$, the list of positions \lst {p^i}{c_i}. First, we turn
this list into a list of strictly positive smaller integers:
\[
p^i_0+1,p^i_1-p^i_0,p^i_2-p^i_1,\ldots,p^i_{c_i-1}-p^i_{c_i-2}.
\]
Consider the concatenation of all sequences above:
\begin{multline}
\label{eq:pos}
p^0_0+1,p^0_1-p^0_0,\ldots,p^0_{c_0-1}-p^0_{c_0-2},\\
p^1_0+1,p^1_1-p^1_0,\ldots,p^1_{c_1-1}-p^1_{c_1-2},\ldots,\\
p^{f-1}_0+1,p^{f-1}_1-p^{f-1}_0,\ldots,p^{f-1}_{c_{f-1}-1}-p^{f-1}_{c_{f-1}-2},
\end{multline}
and store them using the representation for strictly positive numbers. In this
case it is easy to check that the best upper bound is
\begin{equation}
\label{eq:ubpos}
f + \sum_{0\leq i<f} p^i_{c_i-1},
\end{equation}
and the number of elements is the occurrency $g$ of the term. 

We now show how to retrieve the positions of the $i$-th document. Let \lstl sf
be prefix sums of the counts (e.g., $c_i=s_{i+1}-s_i$). We note that the list
provides the starting and ending point of the sequence of positions associated
to a document: the positions of document $i$ occur in~(\ref{eq:pos}) at
positions $j$ satisfying $s_i\leq j<s_{i+1}$. Let \lstl tg be the sequence of prefix sums
of the sequence~(\ref{eq:pos}). It is easy to check that the positions of $i$-th
document can be recovered as follows: \[ p^i_j = t_{s_i+j+1} - t_{s_i} -1 \qquad
0\leq j<c_i. \] We remark that the nice interplay between prefix sums and lists
of natural numbers is essential in making this machinery work: we need the
counts $c_i$ (e.g., to compute a content-based ranking function), but we need
also their prefix sums to locate positions.

% \noindent\textbf{Payloads.} We complete the discussion by showing how to easily
% store fixed-width payloads, such as precomputed impact scores. Such payloads can
% be stored in the upper-bits array of counts or
% positions just after each unary codeword (it is not possible to store
% payloads in the upper-bits array of document pointers
% because it would make skipping impossible). The easy adaptation involves reading
% the payload after each unary-code read, and taking care of compensating for the
% additional space used by payloads when computing the number of zeroes to the left of a position reached through a forward pointer.
 
\section{A Quasi-Succinct BitStream}

We now discuss in detail the bit stream used to store the quasi-succinct
representation described in Section~\ref{sec:qsrep}---in particular, the sizing
of all data involved. 

\begin{figure*}
\centering
\includegraphics{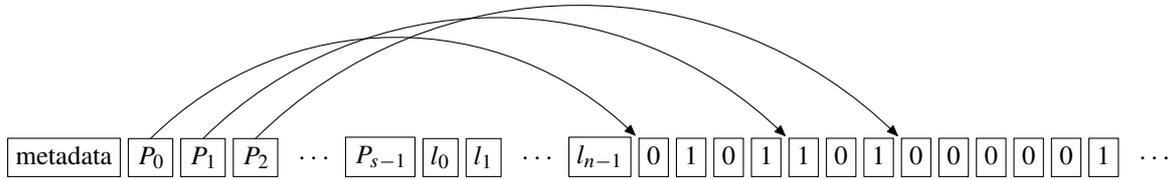}
\caption{\label{fig:bitstream}The bit stream of a quasi-succinct
encoding for a list of $n$ items using $s$ forward pointers. After a self-delimiting metadata section,
there are fixed-width forward pointers, the lower-bits array, and finally
the upper-bits array. In this example, $P_i$ points at the location of the upper-bits array where one would get after $iq$
unary-code reads, with $q=2$. Pointer $P_0$ is never stored explicitly.}
\end{figure*}

Metadata pertaining the whole representation, if present, can be stored
initially in a self-delimiting format. Then, the remaining data is laid out as follows: pointers, lower bits, upper bits
(see Figure~\ref{fig:bitstream}). The rationale behind this layout is that the
upper-bits array is the only part whose length is in
principle unknown: by positioning it at the end of the bitstream, we do not have
to store pointers to the various parts of the stream. The lower-bits array
will be located at position $sw$, where $s$ is the number of pointers and $w$ their
width, and the upper-bits array at position $pw+n\ell$ bits
after the metadata. We can thus compute without further information the starting
point of each part of the stream.

We assume that the number of elements $n$ is known, possibly from the metadata.
The first issue is thus the size and the number of pointers. If the upper bound $u$ is known, we know that
the upper-bits array is $n+\bigl\lfloor
u/2^\ell\bigr\rfloor$ bits long at most, so the width of the pointers is $w=\bigl\lceil\log(n+\bigl\lfloor
u/2^\ell\bigr\rfloor+1)\bigr\rceil$; otherwise, information must be stored in
the metadata part so to be able to compute $w$.

If we are storing forward pointers for unary codes, the number of pointers
will be exactly $\lfloor n/q\rfloor$; otherwise (i.e., if we are storing
forward pointers for negated unary codes), they will be at most
$s=\bigl\lfloor\bigl(n+\bigl\lfloor u/2^\ell\bigr\rfloor\bigr)/
q\bigr\rfloor$.\footnote{We remark that if $u>x_{n-1}$ some of the $s$ pointers might actually be unused. It is sufficient
to set them to zero (no other pointer can be zero) and consider them as
skips to the end of the list.} Again, if the bound $u$ is not known it is
necessary to store information in the metadata part so to be able to compute $s$.

Analogously, if $u$ is not known we need to store
metadata that makes us able to compute $\ell=\lfloor \log(u/n)\rfloor$.

Finally, in the case of a ranked characteristic functions instead of pointers we
store $\lfloor f/q\rfloor$ cumulative ranks of width $w=\lceil \log N\rceil$,
followed by the bitmap representation of the characteristic function.

\section{Laying Out the Index Structure}

We now show how to store in a compact format all metadata that are
necessary to access the lists. For each index component (document pointers,
counts, positions) we write a separate bit stream. We remark that for an index
that provides naturally constant-time access to each element, there
is no point in interleaving data, and this is another advantage of
quasi-succinct encoding, as unnecessary data (e.g., counts and positions for a
Boolean query) need not be examined. As usual, for each term we store three
pointers locating the starting point of the information related to that term in each stream.

The bit stream for \emph{document pointers} contains as metadata the frequency
and the occurrency of the term. We write the occurrency in $\gamma$ code and, if
the occurrency is greater than one, the difference between occurrency and
frequency, again in $\gamma$ code (this ensures that hapaxes use exactly one
bit). This information, together with the number of documents in
the collection, is sufficient to access the quasi-succinct representation of
document pointers (see Section~\ref{sec:repr}).

The bit stream for \emph{counts} contains no metadata. The occurrency
and frequency can be obtained from the pointers stream, and they are sufficient
to access the representation.

The bit stream for \emph{positions} requires to store in the metadata part the
parameter $\ell$ and the skip-pointer size $w$, which we write again in
$\gamma$ code, as the upper bound~(\ref{eq:ubpos}) is not available. Note that if the
occurrency is smaller than $q$, there is no pointer, and in that case we omit
the pointer size. Thus, the overhead for terms with a small number of
occurrences is limited to the parameter $\ell$.\footnote{Actually, it is easy
to check that the overhead for hapaxes is exactly 2 bits with respect to
writing the only existing position in $\delta$ code.}

\section{Implementation Details}

Implementation details are essential in a performance-critical data structure
such as an inverted index. In this section we discuss the main ideas used in our
implementation. While relatively simple, these ideas are essential in obtaining,
besides good compression, a significant performance
increase.

\smallskip\noindent\textbf{Longword addressing.} We either load the index into
memory, or access it as a memory-mapped region. Access happens always by longword, and
shifts are used to extract the relevant data. The bit $k$ of the index is
represented in longword $\lfloor k/64\rfloor$ in position $k\bmod 64$. While
direct access to every point of the bitstream is possible, we keep track of the current
position so that sequential reads use the last longword read as a bit buffer.
Extraction of lower bits requires very few logical operations in
most cases when $\ell$ is small.

\smallskip\noindent\textbf{Reading unary codes.} Reading a unary code is
equivalent to the computation of the least significant bit. We use the beautiful algorithm
based on de Brujin's sequences~\cite{LPRUBSICW}, which is able to locate the
least significant bit using a single multiplication and a table lookup. The
lack of any test makes it a very good choice on superscalar processors, as it
makes prediction and out-of-order execution possible.\footnote{Actually, we
first check whether we can compute the least significant bit using an 8-bit
precomputed table, as the guaranteed high density of the upper bits makes this
approach very efficient.}

Both when looking up an entry and when skipping, we have, however, to perform a
significant number of unary-code reads (on average, $\approx q/2$). To this
purpose, we resort to a \emph{broadword} (a.k.a. \emph{SWAR}, i.e., ``SIMD in A
Register'') \emph{bit search}~\cite{VigBIRSQ}.
The idea is that of computing the number of ones in the current bit buffer using
the classical algorithm for \emph{sideways addition}~\cite{KnuACPBTT}, which
involves few logical operations and a multiplication. If the number of reads we
have to perform exceeds the number of ones in the current buffer, we examine the
next longword, and so on. Once we locate the right longword, we can complete the
search using the broadword \emph{selection} algorithm
presented in~\cite{VigBIRSQ}.

Our experiments show that broadword bit search is extremely effective, unless
the number of reads is very small, as in that case computing iteratively the least
significant bit becomes competitive. Indeed, when skipping a very small number
of position (e.g., less then eight) we simply resort to iterating through the
list.
% \footnote{In case payloads are interleaved with unary
% codes, the bit search algorithm can be adapted to work, albeit the larger the
% payload width, the less efficient the algorithm will become.}

\smallskip\noindent\textbf{Cache the last prefix sum.} When retrieving a
count or the first position of a position list, we have, in theory, to compute two
associated prefix sums. During
sequential scans, however, we can cache the last computed value and use it at
the next call. Thus, in practice, scanning sequentially counts or positions requires just one unary-code
read and one fixed-width bit extraction per item. Reading counts is however made
slower by the necessity to compute the difference between the current and the
previous prefix sum.

\smallskip\noindent\textbf{Trust the processor cache.} The cost of accessing an
in-memory index is largely dominated by cache misses. It is thus not surprising
that using a direct access (i.e., by pointer) can be slower than actually scanning linearly
the upper-bits array using a broadword bit search if our
current position is close to the position to get to. The threshold depend on
architectural issues and must be set experimentally. In our code we use
$q=256$ and we do not use pointers if we can skip to the desired position in
less that $q$ reads.\footnote{Remember, again, that we will actually simulate
such reads using a broadword bit search. } An analogous strategy is used with
ranked characteristic functions: if we have to skip in the vicinity of the
current position and the current index is known we simply read the bitmap, using
the sideways addition algorithm to keep track of the current index.

\section{Experiments}

We have implemented the quasi-succinct index described in the previous
section in Java, and for the part related to document pointers and count, in
C++. All the code used for experiments is available at the MG4J web site. In
this section, we report some experiments that compare its performance against three competitors:
\begin{itemize}
  \item Lucene, a very popular open-source Java search engine (release 3.6.0);
  \item the classical high-performance indices from MG4J~\cite{BoVTREC2005},
  another open-source search engine (release 5.0);
  \item Zettair, a search engine written in C by the Search Engine Group at RMIT
  University.
  \item The Kamikaze\footnote{\texttt{http://sna-projects.com/kamikaze/}}
  library, implementing the PForDelta~\cite{ZHNSSRCCC} sequence compression algorithm (up-to-date repository version from GitHub);
  \item We compare also with recent optimized C code implementing PForDelta compression
  document pointers and count kindly provided by Ding Shuai~\cite{DinPC}.
\end{itemize}
Zettair has been suggested by the TREC organizers as one of the baselines for
the efficiency track. The comparison of a Java engine with a C or C++ engine is
somewhat unfair, but we will see actually
the Java engines turn out to be always significantly faster.

We use several datasets summarized in Table~\ref{tab:datasets}: first, the
classical public TREC GOV2 dataset (about 25 million documents) and a crawl of
around 130 million pages from the \texttt{.uk} domain that is available from the
author. Tokens were defined by transition between alphanumerical to
nonalphanumerical characters or by HTML flow-breaking tags, and they were
stemmed using the Porter2 stemmer\footnote{Zettair, however, supports apparently only
the original Porter stemmer.}. Besides an index considering the whole
HTML document, we created some indices for the title text only (e.g., the
content of the HTML \texttt{TITLE} element), as such indices have significantly
different statistics (e.g., documents are very short).

Additionally, we created a \emph{part-of-speech} index used within the M\'imir
semantic engine~\cite{CTRIESAMPIM}; such indices have a very small number of
terms that represent synctactic elements (nouns, verbs, etc.), very dense
posting lists and a large number of positions per posting: they provide useful
information about the effectiveness of compression when the structure of the
index is not that of a typical web text index. For the same reason, we also
index a collection of about a dozen millions tweets from Twitter.

Small differences in indexing between different search engines are hard to
track: the details of segmentation, HTML parsing, and so on, might introduce
discrepancies. Thus, we performed all our indexing starting from a pre-parsed
stream of UTF-8 text documents. We also checked that the frequency of the
terms we use in our queries is the same---a sanity check showing
that the indexing process is consistent across the engines. Finally, we
checked that the number of results of conjunctive and phrasal queries was consistent across
the different engines, and that \texttt{bpref} scores were in line with those
reported by participants to the Terabyte Track.

\begin{table}
\centering
\begin{tabular}{l|r|r|r|r}
&\multicolumn{1}{c|}{Documents} & \multicolumn{1}{c|}{Terms} &
\multicolumn{1}{c|}{Postings} & \multicolumn{1}{c}{Occurrences}\\
\hline
\multicolumn{5}{c}{TREC GOV2}\\
\hline
Text & 25\,M & 35\,M & 5.5\,G & 23\,G \\
Title &25\,M & $1.1$\,M & 135\,M & 150\,M \\
\hline
\multicolumn{5}{c}{Web \texttt{.uk}}\\
\hline
Text & 130\,M & 99\,M & 21\,G & 62\,G \\
Title& 130\,M & $3.2$\,M & $609$\,M & 691\,M \\
\hline
\multicolumn{5}{c}{M\'imir index}\\
\hline
Token & 1\,M & 49 & $27$\,M & $1.2$\,G \\
\hline
\multicolumn{5}{c}{Tweets}\\
\hline
Text & 13\,M & $2.3$\,M & $147$\,M & $156$\,M \\
\end{tabular}
\caption{\label{tab:datasets}Basic statistics for the datasets used in our
experiments.}
\end{table}

Using MG4J, we have created indices that use $\gamma$ codes for counts, and
either $\delta$ or Golomb codes for pointers and positions\footnote{The Golomb
modulus has been chosen separately for each document. The results we obtain
seems to be within $5$\% of the best compression results obtained
in~\cite{YDSCTPWI}, which suggest a space usage of 21\,MB/query on average for
an average of $20.72$ millions positions per query. A more precise estimate is
impossible, as results in ~\cite{YDSCTPWI} are based on 1000 unknown queries,
and no results about the whole GOV2 collection are provided.}, endowed with a
mild amount of skipping information using around $1\%$ of the index size: we
chose this value because the same amount of space is used by our index to store
forward and skip pointers when $q=256$. These indices (in particular, the ones
based on Golomb codes) are useful to compare compression ratios: if speed is not
a concern, they provide very good compression, and thus they provide a useful
reference points on the compression/speed curve.\footnote{We have also tried
\emph{interpolative coding}~\cite{MoSECIFC}, but on our collections the
difference in compression with Golomb codes was really marginal.}

We remark that we have indexed \emph{every word} of the collections. No stopword
elimination has been applied. Commercial search engines (e.g., Google) are
effortlessly able to search for the phrase ``Romeo and Juliet'', so our engine should
be able to do the same.

\smallskip\noindent\textbf{Compression.} Table~\ref{tab:size} reports a
comparison of the compression ratios. Our
quasi-succinct index compresses always better than $\gamma$/$\delta$, but worse than Golomb codes. In practice,
our index reduces the size of the $\gamma$/$\delta$ index by $\approx10\%$, whereas Golomb
codes reach $\approx20\%$.

\begin{table*}
\centering
\begin{tabular}{l|r|r|r|r|r}
&\multicolumn{1}{c|}{QS} & \multicolumn{1}{c|}{MG4J $\gamma$/$\delta$} & \multicolumn{1}{c|}{Golomb} & \multicolumn{1}{c|}{Lucene}&\multicolumn{1}{c}{Zettair}\\
\hline
\multicolumn{4}{c}{TREC GOV2 (text)}\\
\hline
Pointers  & $7.42$ & $8.47$ & $6.94$&\\ 
Counts    & $2.98$ & $2.56$ & ---& \\
Positions & $10.17$ & $11.11$ & $8.65$ &\\
Overall   & $36.9$\,GB & $40.3$\,GB & $31.9$\,GB & $42.1$\,GB & $40.7$\,GB\\
\hline
\multicolumn{4}{c}{TREC GOV2 (title)}\\
\hline
Pointers  & $10.04$ & $11.44$ & $9.54$& \\ 
Counts    & $1.10$  & $1.14$ & --- &\\
Positions & $3.84$  & $4.63$ & $3.05$& \\
Overall   & $264$\,MB & $308$\,MB & $241$\,MB & 396\,MB & 395\,MB \\
\hline
\multicolumn{4}{c}{Web \texttt{.uk} (text)}\\
\hline
Pointers  & $8.46$ & $9.72$ & $7.98$& \\ 
Counts    & $2.39$ & $2.06$ & --- &\\
Positions & $10.16$ & $10.95$ & $8.41$& \\
Overall   & $108$\,GB & $117$\,GB & $92$\,GB & $126$\,GB\\
\hline
\multicolumn{4}{c}{Web \texttt{.uk} (title)}\\
\hline
Pointers  & $11.75$ & $13.51$ & $11.27$& \\ 
Counts    & $1.13$  & $1.18$ & ---&\\
Positions & $4.36$  & $5.06$ & $3.35$& \\
Overall   & $1.38$\,GB & $1.59$\,GB & $1.26$\,GB & $2.00$\,GB & $2.15$\,GB  \\
\hline
\multicolumn{4}{c}{M\'imir token index}\\
\hline
Pointers  & $1.51$ & $1.42$ & $1.48$& \\ 
Counts    & $6.42$  & $6.28$ & ---&\\
Positions & $5.83$  & $6.22$ & $5.03$& \\
Overall   & $0.96$\,GB & $1.01$\,GB & $0.83$\,GB & $1.34$\,GB & $1.36$\,GB  \\
\hline
\multicolumn{4}{c}{Tweets}\\
\hline
Pointers  & $10.13$ & $10.29$ & $9.22$& \\ 
Counts    & $1.06$  & $1.11$ & ---&\\
Positions & $4.67$  & $5.94$ & $3.86$& \\
Overall   & $302$\,MB & $341$\,MB & $266$\,MB & $423$\,MB & $484$\,MB  \\
\hline
\end{tabular}
\caption{\label{tab:size}A comparison of index sizes. We show the overall index
size, which includes skipping structures, and, if available, the number of bits
per element of each component, excluding skipping structures.}
\end{table*}

The compression of Lucene and Zettair on the text of web pages is not very good
(a $\approx 15\%$ increase w.r.t. our index). This was partially to be expected,
as both Lucene and Zettair use variable-length byte codes for efficiency, and
while such codes are easy to decode, they are ill-suited to compression.
When the distribution of terms and positions is different, however, compression
is significantly worse: for titles we have a $50\%$ increase in size, and for
the M\'imir semantic index or tweets a $40\%$ increase.
This is somewhat typical: variable-length byte codes compress most positions in
a single byte if the distribution of words comes from a ``natural'' distribution
on documents of a few thousand words.
Using shorter documents (e.g., titles and tweets) or a different distribution
(e.g., a semantic index) yields very bad results. A $50\%$ increase in
size, indeed, can make a difference.

While we are not aiming at the best possible compression, but rather at high
speed, it is anyway relieving to know that we are improving (as we shall see
shortly) \emph{both} compression \emph{and} speed with respect to these engines.

Interestingly, counts are the only index component for which
we obtain sometimes worse results than $\gamma$ coding. This is somewhat to be
expected, as we are actually storing their prefix sums. The impact of counts on
the overall index, however, is quite minor, as shown by the small final index
size.

\smallskip\noindent\textbf{Speed.} Benchmarking a search engine brings up
several complex issues. In general, the final answer is bound to the
architecture on which the tests were run, and on the type of queries. A definite
answer can be given only against a real workload.\footnote{Note that in
real-world search engines the queries that are actually solved are very
different by those input by the user, as they undergo a number of rewritings. As
a consequence, blindingly analzying queries from large query logs in disjunctive
or conjunctive mode cannot give a reliable estimate the actual performance of an
index.} Our tests were performed on a recent workstation sporting a 3.4\,GHz
Intel i7-3770 CPU with 8\,MiB of cache and 16\,GiB of RAM.

We aim at comparing speed of in-memory indices, as one of the main reasons to
obtain smaller indices is to make more information fit into memory; moreover,
the diffusion of solid-state disks makes this approach reasonable. Thus, in our
tests we resolve each query three times before taking measurements. In this way
we guarantee that the relevant parts of the index have been actually read and
memory mapped (for MG4J and Lucene, or at least cached by the file system, for
Zettair), and we also make sure that the Java virtual machine is warmed up
and has performed inlining and other runtime optimizations. With this setup, our
tests are highly repeatable and indeed the relative standard deviation over
several runs is less than 3\%.

We used the 150 TREC Terabyte track (2004$-$2006) title queries in conjunctive,
phrasal and proximity form (in the latter case, the terms in the
query must appear in some order within a window of 16 words).
We also extracted the terms appearing in the queries and used them as queries to
measure pure scanning speed: all in all, we generated 860 queries. MG4J and
Lucene were set up to compute the query results without applying any ranking
function. Zettair was set to Okapi BM25 ranking~\cite{SWRPMIRI}, which appeared
to have the smaller impact on the query resolution time (no ``no-ranking'' mode
is available).

% There is however an important difference between Zettair and Lucene or MG4J:
% Zettair can only parse and score ``flat'' queries (conjunctive, disjunctive or
% phrasal). Thus, its query resolution logic is significantly simpler than
% that of Lucene or MG4J, which are able to score arbitrary structured queries.

All engines were set up to return a single result, so that the logic needed to
keep track of a large result size would not interfere with the evaluation. The
results are shown in Table~\ref{tab:trecmem}.
The first column (QS) shows the results of query resolution on a quasi-succinct
index. The third column (MG4J) for a $\gamma$/$\delta$-coded high-performance
MG4J index. The fourth column for Lucene, and the last column for Zettair.

The second column (QS*) needs some explanation. Both Lucene and MG4J interleave
document pointers and counts. As a consequence, resolving a pure Boolean query
has a higher cost (as counts are read even if they are not necessary), but
ranked queries require less memory/disk access. To simulate a similar behaviour
in our setting, we modified our code so to force it to read the count of every
returned document pointer. This setting is of course artificial, but it provides
a good indication of the costs of iterating and applying a count-based ranking
function, and it will be the based of our comparison. For phrasal and proximity
queries there is no difference between QS and QS* as counts have in any case to
be read to access positions.

First of all, we note that decoding a quasi-succinct index is slightly ($\approx
7$\%) faster than decoding a gap-compressed index that uses variable-byte codes.
It is nonetheless important to notice that our timings for purely boolean
resolution (QS) are much lower, and this can be significant in a complex query
(e.g., a conjunction of disjunctively expanded terms). Zettair is much slower.

More interestingly, we have a $\approx 50$\% improvement for
conjunctive queries, a $\approx 40$\% improvement for phrasal queries and a
$\approx 60$\% improvement for proximity queries:
being able to address in average constant time every element of the index 
is a real advantage. We also remind the reader that we are comparing a
Java prototype with a mature implementation.

We expect the asymptotic advantage of quasi-succinct indices to be
more evident as the collection size grows. To test this hypothesis, we performed
further experiments using the Web \texttt{.uk} collection and 1000 multi-term queries randomly selected
from a large search-engine query log. The results are shown in
Table~\ref{tab:uk}: now conjunctive and proximity queries are more thrice
faster with respect to Lucene.

In Table~\ref{tab:p4d} we show some data comparing in-memory quasi-succinct
indices with PForDelta code. The data we display is constrained by some
limitations: the Kamikaze library does not provide count storage; and the
optimized C code we are using~\cite{DinPC} does not provide positions. This is
an important detail, as quasi-succinct indices trade some additional efforts in decoding counts
(i.e., computing their prefix sums) in exchange for constant-time access to
positions. Our main goal is to speed up positional access---indeed, nothing
prevents using PForDelta for document pointers and storing counts and positions
as described in this paper (or even using a separate PForDelta index without
positions as a first-pass index).

Kamikaze turns out to be slightly slower for scanning term lists, and almost
twice as slow when computing conjunctive queries.
To estimate the difference in compression, we computed
the space used by the document pointers of our TREC collections using Kamikaze:
the result is an increase of $\approx55$\% in space usage. While not extremely
relevant for the index size (positions are responsible mostly for the size of an
index), it shows that we would gain no advantage from storing pointers using
PForDelta in a Java engine.\footnote{Note that storing \emph{positions} with
PForDelta codes is known to give a compression rate close to that provided by
variable-byte coding~\cite{YDSCTPWI}.}

The comparison of C implementations, on the other hand, is definitely in favour
of PForDelta: apart from pointer enumeration our C implementation is
slower, in particular when enumerating terms and their counts.

There are some important \textit{caveats}, however: the code we have been
provided for PForDelta testing~\cite{DinPC} is a bare-bone, heavily
optimised C benchmarking implementation that is able to handle only 32-bit
document pointers and has a number of limitations such as hardwired constants
(e.g., the code needs to be recompiled if the number of document in the
collection changes). Our C++ code is a 64-bit fully usable implementation
derived from a line-by-line translation of our Java prototype code that could be
certainly improved by applying CPU-conscious optimizations.
A more realistic comparison would require a real search engine using PForDelta
to solve queries requiring positional information, it happens in
Table~\ref{tab:trecmem}.\footnote{Such an engine is not available, to the best
of the authors's knowledge. The authors of~\cite{YDSCTPWI} have refused
to make their engine available.}

In Table~\ref{tab:uk}
we report similar data for our Web \texttt{.uk} collection: also in this case, a
larger collection improves our results (in particular, conjunctive queries are
only $\approx13$\% slower than PForDelta, instead of $\approx 21$\%).

\begin{table}
\centering
\begin{tabular}{l|r|r|r|r|r|}
&\multicolumn{1}{c|}{QS}&\multicolumn{1}{c|}{QS*}
&\multicolumn{1}{c|}{MG4J} &
\multicolumn{1}{c|}{Lucene}&\multicolumn{1}{c}{Zettair}\\
\hline
Terms   & $4.51$ & $7.82$ & $10.33$ & $8.26$ & $19.17$\\
And     & $1.29$ & $1.79$ & $4.90$ & $3.90$ & $20.92$ \\ 
Phrase  & $4.00$ & --- & $11.01$ & $6.77$ & $21.14$\\
Proximity & $4.76$ & --- & $12.15$ & $12.05$ & ---\\
\hline
\end{tabular}
\caption{\label{tab:trecmem} Timings in seconds for running the test queries
from the TREC Terabyte track on GOV2 without scoring. The column QS
shows the timings for resolving a query on a quasi-succinct
indices, whereas the column QS* shows the timings for a modified
version in which counts are forced to be read for each decoded document pointer. 
Measurements were taken after three executions of each query, with memory map and disk 
caches already filled. Note that Zettair is actually reading from disk and scoring the queries, 
whereas in the other cases pointers and counts are being read from a memory-mapped region and no score is being computed.}
\end{table}

\begin{table*}
\centering
\begin{tabular}{l|r|r|r|r|r|r|r|}
&\multicolumn{1}{c|}{QS}&\multicolumn{1}{c|}{QS*}&\multicolumn{1}{c|}{Kamikaze}&\multicolumn{1}{c|}{QS(C)}&\multicolumn{1}{c|}{QS*(C)}&\multicolumn{1}{c|}{PFD(C)}&\multicolumn{1}{c|}{PFD*(C)}\\
\hline
Terms   & $3.83$ & $7.30$ & $4.23$ & $1.61$ & $4.05$ & $1.57$ & $2.39$  \\
And     & $1.16$ & $1.62$ & $2.08$ & $0.91$ & $1.25$ & $0.75$ & $0.87$ \\ 
\hline
\end{tabular}
\caption{\label{tab:p4d} Timings in seconds for running the term and
conjunctive test queries from the TREC Terabyte
track on GOV2 directly from RAM. Timings for quasi-succinct indices are
provided both for Java and C++ 64-bit implementations. PForDelta timings have
been computed both using the Kamikaze library and using optimized 32-bit C code
provided by Ding Shuai~\cite{DinPC}.
Starred versions include reading counts for all returned document pointers.}
\end{table*}

\begin{table}
\centering
\begin{tabular}{l|r|r|r|}
&\multicolumn{1}{c|}{QS}&\multicolumn{1}{c|}{QS*}&\multicolumn{1}{c|}{Lucene}\\
\hline
Terms     & $70.9$  & $132.1$ & $130.6$\\
And       & $27.5$  &  $36.7$ & $108.8$\\ 
Phrase    & $78.2$  & ---  & $127.2$ \\
Proximity & $106.5$ & ---  & $347.6$ \\
\hline
\end{tabular}
\caption{\label{tab:uk} Timings in seconds for running 1000 randomly
selected queries from a search-engine query log on the
Web \texttt{.uk} collection. See also Table~\ref{tab:trecmem}.}
\end{table}

\begin{table}
\centering
\begin{tabular}{l|r|r|r|r|}
&\multicolumn{1}{c|}{QS(C)}&\multicolumn{1}{c|}{QS*(C)}&\multicolumn{1}{c|}{PFD(C)}&\multicolumn{1}{c|}{PFD*(C)}\\
\hline
Terms      & $23.8$ & $56.8$ & $23.6$ & $31.6$ \\
And        & $19.2$ & $24.5$ & $16.9$ & $19.4$\\ 
\hline
\end{tabular}
\caption{\label{tab:uk} Timings in seconds for running 1000 randomly
selected queries from a search-engine query log on the
Web \texttt{.uk} collection. See also Table~\ref{tab:p4d}.}
\end{table}

\section{Some anecdotal evidence}

While running several queries in controlled conditions is a standard practice,
provides replicable results and gives a general feeling of what is
happening, we would like to discuss the result of a few selected queries that highlight the strong points of
our quasi-succinct indices. We keep the same settings as in the previous section
(e.g., ranked queries repeated several times to let the cache do its work). All
timings are in milliseconds.

\smallskip\noindent\textbf{Dense terms.} We start by enumerating all documents
in which the term ``and'' appears ($\approx18$ millions):
\begin{center}
\begin{tabular}{r|r|r|r|r|r|}
\multicolumn{1}{c|}{QS}&\multicolumn{1}{c|}{Kamikaze}&\multicolumn{1}{c|}{QS*}
&\multicolumn{1}{c|}{MG4J} & \multicolumn{1}{c|}{Lucene}&\multicolumn{1}{c}{Zettair}\\
\hline
$72.4$ & $179.2$ & $234.6$ & $488.5$ & $283.6$ & $1246.5$\\
\hline
\end{tabular}
\end{center}
In this case, our quasi-succinct index is a ranked characteristic function.
Reads are particularly fast (just a unary code read), and combined with count
reading faster than Lucene. Note
that we compress this list at $\approx 1.38$ bits per pointer, against the
$\approx 2.38$ bits of Kamikaze and the $8$ bits of Lucene. The slowness of
Zettair is probably due to the fact that 
positional information is interleaved with document pointers, so it is necessary
to to skip over it. 

Another example (this time using an Elias--Fano representation) is
the enumeration of all documents in which the term ``house'' appears ($\approx 2$
millions):
\begin{center}
\begin{tabular}{r|r|r|r|r|r|}
\multicolumn{1}{c|}{QS}&\multicolumn{1}{c|}{Kamikaze}&\multicolumn{1}{c|}{QS*}
&\multicolumn{1}{c|}{MG4J} & \multicolumn{1}{c|}{Lucene}&\multicolumn{1}{c}{Zettair}\\
\hline
$17.2$ & $19.4$ & $31.9$ & $42.2$ & $33.2$ & $69.1$\\
\hline
\end{tabular}
\end{center}
An Elias--Fano list requires recovering also the lower bits, and thus it is
slightly slower: overall, if we read counts we are just slightly faster than
Lucene, as expected.

\smallskip\noindent\textbf{Conjunction of correlated terms.} Consider the conjunction of the 
terms ``home'' and ``page'', which appears in about one fifth of the
documents in the GOV2 collection:
\begin{center}
\begin{tabular}{r|r|r|r|r|r|}
\multicolumn{1}{c|}{QS}&\multicolumn{1}{c|}{Kamikaze}&\multicolumn{1}{c|}{QS*}
&\multicolumn{1}{c|}{MG4J} & \multicolumn{1}{c|}{Lucene}&\multicolumn{1}{c}{Zettair}\\
\hline
$204$ & $295$ & $420$ & $561$ & $416$ & $933$\\
\hline
\end{tabular}
\end{center}
We can see that in this case quasi-succinct indices are already better than
Kamikaze at conjunction, but nonetheless the high correlation makes our
constant-time skipping not so useful.

% Timings for the \emph{phrasal} query ``home
% page'' confirm this fact:
% \begin{center}
% \begin{tabular}{r|r|r|r|r|}
% \multicolumn{1}{c|}{QS}&\multicolumn{1}{c|}{MG4J} & \multicolumn{1}{c|}{Lucene}&\multicolumn{1}{c}{Zettair}\\
% \hline
% $2623$ & $3155$ & $2247$ & $2053$\\
% \hline
% \end{tabular}
% \end{center}
% Now the engines have essentially to read wholly all posting lists. No skipping
% is possible (it would be actually detrimental). Most of the time is spent trying to 
% figure out which of the documents containing the two terms actually contains the
% two terms in a row---and here Java's overhead becomes visible.

On the other hand, consider the conjunction of the 
terms ``good'', ``home'' and ``page'', which appears in about 1/30th of
the documents in the GOV2 collection:
\begin{center}
\begin{tabular}{r|r|r|r|r|r|}
\multicolumn{1}{c|}{QS}&\multicolumn{1}{c|}{Kamikaze}&\multicolumn{1}{c|}{QS*}
&\multicolumn{1}{c|}{MG4J} & \multicolumn{1}{c|}{Lucene}&\multicolumn{1}{c}{Zettair}\\
\hline
$73$ & $153$ & $164$ & $471$ & $294$ & $709$\\
\hline
\end{tabular}
\end{center}
The query is now more complex, but, more importantly, there is a term that is
significantly less frequent than the other two. Quasi-succinct indices have now
a significant advantage.

It is interesting to compare the above table with the
timings for the \emph{phrasal} query ``home page'':
\begin{center}
\begin{tabular}{r|r|r|r|r|}
\multicolumn{1}{c|}{QS}&\multicolumn{1}{c|}{MG4J} & \multicolumn{1}{c|}{Lucene}&\multicolumn{1}{c}{Zettair}\\
\hline
$1282$ & $1693$ & $1228$ & $977$\\
\hline
\end{tabular}
\end{center}
Now the engines have essentially to read wholly all posting lists. No skipping
is possible (it would be actually detrimental). Most of the time is spent trying to 
figure out which of the documents containing the three terms actually contains
the three terms in a row. The overhead of Java becomes here 
visible---this is indeed our only example in which Zettair is the fastest
engine.

It is interesting to compare the above table with the
timings for the phrasal query ``good home page'':
\begin{center}
\begin{tabular}{r|r|r|r|r|}
\multicolumn{1}{c|}{QS}&\multicolumn{1}{c|}{MG4J} & \multicolumn{1}{c|}{Lucene}&\multicolumn{1}{c}{Zettair}\\
\hline
$540$ & $1251$ & $880$ & $795$\\
\hline
\end{tabular}
\end{center}

\smallskip\noindent\textbf{Conjunction of uncorrelated terms.} The terms
``foo'' and ``bar'' appear in about 650\,000 documents, but they
co-occur just in about 5\,000:
\begin{center}
\begin{tabular}{r|r|r|r|r|r|}
\multicolumn{1}{c|}{QS}&\multicolumn{1}{c|}{Kamikaze}&\multicolumn{1}{c|}{QS*}
&\multicolumn{1}{c|}{MG4J} & \multicolumn{1}{c|}{Lucene}&\multicolumn{1}{c}{Zettair}\\
\hline
$1.27$ & $2.28$ & $2.00$ & $7.09$ & $3.11$ & $35.39$\\
\hline
\end{tabular}
\end{center}
The smallness of the intersection gives to our skipping logic a greater
advantage than in the previous case.

The terms ``fast'' and ``slow'' appear in about 1,000\,000 documents, but they
co-occur just in about 50\,000:
\begin{center}
\begin{tabular}{r|r|r|r|r|r|}
\multicolumn{1}{c|}{QS}&\multicolumn{1}{c|}{Kamikaze}&\multicolumn{1}{c|}{QS*}
&\multicolumn{1}{c|}{MG4J} & \multicolumn{1}{c|}{Lucene}&\multicolumn{1}{c}{Zettair}\\
\hline
$9.21$ & $10.0$ & $12.45$ & $25.21$ & $17.20$ & $45.22$\\
\hline
\end{tabular}
\end{center}

\smallskip\noindent\textbf{Complex selective queries.} The
query ``foo bar fast slow'' has $\approx250$ results:
\begin{center}
\begin{tabular}{r|r|r|r|r|r|}
\multicolumn{1}{c|}{QS}&\multicolumn{1}{c|}{Kamikaze}&\multicolumn{1}{c|}{QS*}
&\multicolumn{1}{c|}{MG4J} & \multicolumn{1}{c|}{Lucene}&\multicolumn{1}{c}{Zettair}\\
\hline
$1.25$ & $2.20$ & $1.32$ & $7.21$ & $7.48$ & $68.26$\\
\hline
\end{tabular}
\end{center}
The more the query becomes selective, the greater the advantage of average
constant-time positioning. Note, in particular, that the timing for QS*
\emph{decreases}, as less counts have to be retrieved (and they can be
retrieved quickly).

\smallskip\noindent\textbf{Phrases with stopwords.} As we remarked in
the previous section, we should be able to search for the exact phrase ``Romeo and Juliet'':
\begin{center}
\begin{tabular}{r|r|r|r|r|}
\multicolumn{1}{c|}{QS}&\multicolumn{1}{c|}{MG4J} & \multicolumn{1}{c|}{Lucene}&\multicolumn{1}{c}{Zettair}\\
\hline
$2.53$ & $15.12$ & $6.36$ & $1203.85$\\
\hline
\end{tabular}
\end{center}
Zettair performs particularly badly in this case. Our ability
to address quickly any position of the index more than doubles the speed of our
answer with respect to Lucene. This can be seen also  from
the timings for the conjunctive query containing ``Romeo'', ``and'', and ``Juliet'':
\begin{center}
\begin{tabular}{r|r|r|r|r|r|}
\multicolumn{1}{c|}{QS}&\multicolumn{1}{c|}{Kamikaze}&\multicolumn{1}{c|}{QS*}
&\multicolumn{1}{c|}{MG4J} & \multicolumn{1}{c|}{Lucene}&\multicolumn{1}{c}{Zettair}\\
\hline
$0.51$ & $2.47$ & $0.92$ & $6.64$ & $3.41$ & $1244.03$\\
\hline
\end{tabular}
\end{center}
The number of results increases by $\approx 15\%$.

\smallskip\noindent\textbf{Proximity.} As Table~\ref{tab:trecmem} shows,
quick access to positions improve significantly another important aspect of a search engine: proximity
queries. Here we show a roundup of the previous conjunctive queries resolved
within a window of 16 words:
\begin{center}
\begin{tabular}{l|r|r|r|r|}
&\multicolumn{1}{c|}{QS}&\multicolumn{1}{c|}{MG4J} &
\multicolumn{1}{c|}{Lucene}\\
\hline
home page & $1625.30$ & $2134.45$ & $2079.52$ \\
good home page & $754.25$ & $1498.17$ & $1203.64$ \\
foo bar &  $3.22$ & $12.84$ & $7.40$\\
fast slow & $23.33$ & $50.68$ & $39.11$\\
foo bar fast slow & $1.48$ & $9.15$ & $12.40$ \\
romeo and juliet &$3.22$& $16.20$ & $11.41$ \\
\hline
\end{tabular}
\end{center}
These results show, in particular, that quick access to positions makes
proximity computation always \emph{faster} for more complex queries.

\smallskip\noindent\textbf{C implementation.} Finally, we show the timings for
the same set of queries using C implementations of PForDelta and quasi-succinct
indices:
\begin{center}
\begin{tabular}{l|r|r|r|r|}
&\multicolumn{1}{c|}{QS(C)}&\multicolumn{1}{c|}{PFD(C)}&\multicolumn{1}{c|}{QS*(C)}&\multicolumn{1}{c|}{PFD*(C)}\\
\hline
home page & $159.08$ & $134.14$ & $316.91$ & $162.40$ \\
good home page & $63.06$ & $67.71$ & $121.36$ & $84.18$ \\
foo bar & $0.73$ & $0.67$ & $1.01$ & $0.83$\\
fast slow & $6.34$ &$4.42$ & $8.36$ &$5.04$ \\
foo bar fast slow & $0.74$ & $0.74$ & $0.82$ & $0.79$ \\
romeo and juliet & $0.29$ & $0.90$ & $0.56$ & $1.00$ \\
\hline
\end{tabular}
\end{center}
We already know from Table~\ref{tab:p4d} that PForDelta optimised code is
significantly faster at retrieving counts (see columns QS*(C) and PDF*(C));
the same comments apply. As expected, albeit in general slower our
quasi-succinct C++ implementation is faster at solving queries with a
mix of high-density and low-density terms (``good home page'' and ``romeo and juliet'').

\section{Conclusions}

We have presented a new inverted index based on the quasi-succinct encoding of
monotone sequences introduced by Elias and on ranked characteristic functions.
The new index provides better compression than typical gap-encoded indices, with
the exception of extremely compression-oriented techniques such as Golomb or
interpolative coding. When compared with indices based on gap compression using
variable-length byte encoding (Lucene) or $\gamma$/$\delta$ codes (MG4J), not
only we provide better compression, but significant speed improvement over
conjunctive, phrasal and proximity queries. In general, any search engine
accessing positional information for selecting or ranking documents out of a
large collection would benefit from quasi-succinct indices (an example being
tagged text stored in parallel indices).

Our comparison with a C implementation of PForDelta compression for pointers and
counts showed that PForDelta is slightly faster than quasi-succinct indices in
computing conjunction, and significantly faster at retrieving counts, albeit in
queries mixing terms with high and low frequency quasi-succinct indices can be
extremely faster. Moreover, PForDelta (more precisely, the Kamikaze library) use
$55$\% more space than a quasi-succinct index to compress pointers from the GOV2
collection.

A drawback of quasi-succinct indices is that some basic statistics (in
particular, frequency, occurrency and the bound~(\ref{eq:ubpos})) must be known
before the index is built. This implies that to create a quasi-succinct
index from scratch it is necessary to temporary cache in turn each posting
list (e.g., using a traditional gap-compressed format) and convert it to
the actual encoding only when all postings have been generated. While it is easy
to do such a caching offline, it could slow down index construction.

On the other hand, this is not a serious problem: in practice, large indices
are built by scanning incrementally (possibly in parallel) a collection, and merges
are performed periodically over the resulting \emph{segments} (also called
\emph{barrels} or \emph{batches}). Since during the construction of a segment it
is trivial to store the pieces of information that are needed to build a
quasi-succinct index, there is no need for an actual two-pass construction:
segments can be compressed using gap encoding, whereas large indices can
be built by merging in a quasi-succinct format.

% We created Javaprototype for ease of development and testing, but clearly
% there are several implementation details that can be improved, leading to even
% better performance. A particularly well-behaved property of our index, indeed,
% is that decoding an element (whether a document pointer, a count or a position)
% requires ``glueing'' two pieces of information \emph{that can be retrieved
% independently}. Indeed, we believe that one of the reason for the high speed of
% raw decoding in C is that operations to retrieve data from the upper-bits and
% the lower-bits array can be executed out-of-order and with a high degree of
% parallelism---there is single serialization point which is the or operation that
% combines upper and lower bits. This property might suggest further optimizations
% of the scanning process.

Note that if computing the least significant bit, selection-in-a-word and
sideways addition were available in \emph{hardware}, the decoding speed of a
quasi-succinct index would significantly increase, as about 30\% decoding time
is spent reading unary codes. It is difficult to predict the impact of such
hardware instructions on skipping, but we would certainly expect major speedups.
In Java virtual machines, this would lead a to better \emph{intrinsification} of
methods such as \texttt{Long.numberOfTrailingZeros()}, whereas the \texttt{gcc}
compiler could provide faster versions of built-in functions such as
\verb|__builtin_ctzll()|.

% We want also remark that the possibility of considering \emph{every} part of the
% index randomly accessible paves the way to several changes in the way, for
% instance, positions are handled. In the case of indices with several positions
% per postings (e.g., tagged text), it might be interesting to explore the
% possibility of expanding positions lazily as the caller requires; in some cases
% (e.g., phrase algorithms) instead of enumerating all positions the caller might
% invoke a ``skip beyond position $p$'' primitive that would be trivial to
% implement quickly by pure SWAR bit search (storing additional skip
% pointers is also possible, but it would increase the index size, as it would
% require to store positions using the standard representation).

An interesting area of future research would be extending the techniques
described in this paper to \emph{impact-sorted} indices, in which documents are
sorted following a retrieval-based \emph{impact order}~\cite{AnMPQEUPCI}, and
only documents pointers with the same impact are monotonically increasing. A
technique similar to that used in this paper to store positions (i.e., a
different encoding for the start of each block) might provide new interesting
tradeoffs between compression and efficiency.

% We leave for future work the comparison with gap-compressed index using codes
% strictly conceived for quick decoding~\cite{AnMIICUWABC,ZHNSSRCCC}, which lie at
% the other extreme of the compression/speed curve with respect to Golomb codes.

\section{Acknowledgments}

The author would like to thank Roi Blanco for an uncountable number of useful
suggestions and for moral support.

\bibliography{biblio}

\end{document}